\documentclass[acmsmall]{acmart}

\def\BibTeX{{\rm B\kern-.05em{\sc i\kern-.025em b}\kern-.08emT\kern-.1667em\lower.7ex\hbox{E}\kern-.125emX}}
\usepackage{xspace}
\usepackage{multirow}
\usepackage{xcolor}
\usepackage{subcaption}
\usepackage{caption}
\usepackage{array}
\usepackage{booktabs} %
\usepackage{amsmath}
\usepackage{multirow}
\usepackage{wasysym}
\usepackage{tikz}
\usetikzlibrary{shapes}
\usepackage{footmisc}

\usepackage[final]{changes}

\setcopyright{acmcopyright}
\acmJournal{PACMHCI}
\acmYear{2019} \acmVolume{3} \acmNumber{CSCW} \acmArticle{163} \acmMonth{11} 
\acmArticle{163} \acmMonth{11} \acmPrice{15.00}
\acmDOI{10.1145/3359265}

\received{April 2019} 
\received[revised]{June 2019}
\received[accepted]{August 2019}

\begin{document}

\newcommand{\titletext}{Content Removal as a Moderation Strategy:
Compliance and Other
Outcomes in the ChangeMyView Community
}

\title{\titletext}
\author{Kumar Bhargav Srinivasan}
\email{kumar.srinivasan@colorado.edu}
\affiliation{%
  \institution{University of Colorado Boulder, USA}
  \city{Boulder}
  \state{Colorado}
  \postcode{80309}
}
\author{Cristian Danescu-Niculescu-Mizil }
\email{cristian@cs.cornell.edu}
\affiliation{%
  \institution{Cornell University, USA}
  \city{Ithaca}
  \state{New York}
  \postcode{14850}
}
\author{Lillian Lee}
\orcid{0000-0003-4770-1712} %
\email{llee@cs.cornell.edu} %
\affiliation{%
  \institution{Cornell University, USA}
  \city{Ithaca}
  \state{New York}
  \postcode{14850}
}
\author{Chenhao Tan}
\authornote{Principal contact author.}
\email{chenhao@chenhaot.com}
\affiliation{%
  \institution{University of Colorado Boulder, USA}
  \city{Boulder}
  \state{Colorado}
  \postcode{80309}
}

\thanks{We thank the anonymous reviewers and chairs for their very thoughtful feedback; their advice and questions resulted in substantial improvements to this paper.
We also thank the members of the NLP+CSS research group at CU Boulder,
Liye Fu, Bobby Kleinberg,
and Jon Kleinberg for their insightful comments and discussions; and
Kal Turnbull and the CMV moderators for providing access to moderation data.
This material is based upon work supported by the National Science Foundation under
grants
BIGDATA SES-1741441 and CAREER IIS-1750615. Any opinions, findings, and conclusions or recommendations expressed in this material are those of the authors and do not necessarily reflect the views of the National Science Foundation}
\renewcommand{\shortauthors}{Kumar Bhargav Srinivasan et al.}
\newif\ifshowcomments
\showcommentstrue
\newcommand{\chenhao}[1]{\textbf{\color{blue}[** #1 **]}}
\newcommand{\kumar}[1]{\textbf{\color{red}[** #1 **]}}
\newcommand{\llee}[1]{\textbf{\color{magenta}[** #1 **]}}
\newcommand{\cd}[1]{\textbf{\color{green}[** #1 **]}}
\newcommand{\para}[1]{\noindent{\bf #1}\xspace}
\ifshowcomments
\else
\renewcommand{\chenhao}[1]{}
\renewcommand{\kumar}[1]{}
\renewcommand{\llee}[1]{}
\renewcommand{\cd}[1]{}
\fi

\newcommand{\secref}[1]{Section~\ref{#1}\xspace} %
\newcommand{\figref}[1]{Figure~\ref{#1}\xspace}
\newcommand{\tableref}[1]{Table~\ref{#1}\xspace}
\newcommand{\communityname}[1]{{\sf #1}\xspace}
\newcommand{\MYhref}[3][blue]{\href{#2}{\color{#1}{#3}}}
\newcommand{\posttree}{post tree\xspace} %
\newcommand{\posttrees}{{\posttree}s\xspace}
\newcommand{\Posttrees}{Post trees\xspace}
\newcommand{\PostTrees}{Post Trees\xspace}
\newcommand{\punishee}{affected individual\xspace}
\newcommand{\Punishee}{Affected individual\xspace}
\newcommand{\punishees}{affected individuals\xspace}
\newcommand{\Punishees}{Affected individuals\xspace}
\newcommand{\punisheespos}{affected individuals'\xspace}
\newcommand{\Punisheespos}{Affected individuals'\xspace}
\newcommand{\novelexperimentalsetup}{a novel experimental setup\xspace}
\newcommand{\methodology}{experimental setup\xspace}
\newcommand{\delayedreaction}{delayed feedback\xspace}
\newcommand{\Delayedreaction}{Delayed feedback\xspace}
\newcommand{\background}{background\xspace}
\newcommand{\treatment}{affected\xspace}
\newcommand{\treatmentposttrees}{affected trees\xspace}
\newcommand{\nontreatmentposttrees}{non-affected trees\xspace}
\newcommand{\treatmentposttree}{affected tree\xspace}
\newcommand{\nontreatmentposttree}{non-affected tree\xspace}
\newcommand{\Treatment}{Affected\xspace}
\newcommand{\nontreatment}{non-affected\xspace}
\newcommand{\Nontreatment}{Non-affected\xspace}
\newcommand{\preremoval}{pre-removal\xspace}
\newcommand{\Preremoval}{Pre-removal\xspace}

\newcommand{\cut}[1]{}
\newcommand{\addterm}[2]{\textbf{#1}\space#2\\}
\newcommand{\introterm}[1]{{\em {\bf #1}}\xspace}
\newcommand{\figuredir}{figures/}
\newcommand{\addFigure}[2]{\includegraphics[width={#1}]{{\figuredir}/{#2}}}
\newcommand{\citeposs}[1]{\citet{#1}'s }


\begin{abstract}
Moderators of online communities often employ comment deletion as a tool.
We ask here whether, beyond the positive effects of shielding a community from
undesirable content, does comment removal actually cause the behavior of the
comment's author to improve?
We examine this question in a particularly well-moderated community, the
ChangeMyView subreddit.

The standard analytic approach of interrupted time-series analysis unfortunately
cannot answer this question of causality because it
fails to distinguish the effect of {\em having made} a non-compliant comment from the
effect of
{\em being subjected to moderator removal} of that comment.
We
therefore leverage a 
``\delayedreaction'' approach based on the observation that
some
users may remain active between
the time when they posted the non-compliant comment and the time when that comment is deleted.
Applying this approach to such users,
we reveal
the causal role of comment deletion in reducing
immediate
noncompliance rates,
although
we do not find evidence of it having a
causal role
in inducing other behavior improvements.
Our work thus empirically demonstrates both the promise
and some potential limits
of content removal as a
positive
moderation strategy,
and points to future directions for identifying causal effects from observational data.
\end{abstract}


\maketitle

\section{Introduction}
\label{sec:intro}
Moderators of online communities often remove inappropriate content in order to shield users
from undesirable material
 and enforce community rules.
But
do authors of a deleted comment alter their behavior, for better or perhaps even for worse, after their comment is removed?
In this work, we
begin exploring
this question by studying
a particularly well-moderated
and goal-oriented
community, ChangeMyView.

\urldef\urlmedia\url{https://changemyview.net/subreddit/#media-coverage}

The ChangeMyView (CMV) subreddit\footnote{\url{https://www.reddit.com/r/changemyview/}.
Recently, a spinoff website, \url{https://changeaview.com}, was created
along roughly the same lines.
}
hosts conversations where someone expresses a view,  typically a controversial opinion,
with the intention of hearing alternate perspectives;
others then try to change that
person's mind. Despite being fundamentally based
on
people arguing with each other,
ChangeMyView has a
reputation\footnote{Media coverage listed at \urlmedia.} for being remarkably civil and productive for an
open internet site; here is what one journalist who joined it said:
\begin{quote}
My opinion was no longer a ``take'' fitted to Twitter or an op-ed. It was a responsible perspective, honed in a collegial atmosphere. ....
In a culture of brittle talking points that we guard with our lives, Change My View is a source of motion and surprise. Who knew that my most heartening ideological conversation in ages would involve gonads, gender wars, and for heaven's sake Reddit? \cite{heffernan:2018a}
\end{quote}

Users consider moderator intervention to be one of the key factors behind the
quality of discussion in CMV \cite{jhaver_designing_2017,heffernan:2018a}, and
comment removal is the moderators'\footnote{Non-moderators
may downvote or report other people's content, but can't remove it.}
front-line strategy. (Removal is considered a public formal warning, which affected
users can appeal; users are
typically only banned from the site after
3 comment removals.)
We aim to understand the effects of the considerable
effort devoted by moderators to this activity (they manually removed 22,788  comments between January 2015 and March 2018)
on the
subsequent behavior of a
deleted comment's author, or {\em \punishee}.
\added{
There is significant subtlety, discussed in
Section \ref{sec:methods},
in formulating concrete hypotheses that are
testable with data currently available, that is, do not require changes to the
moderators' modus operandi.
Working under these constraints, we formulate a
set of hypotheses taking the following template form:
}
\begin{quote}
\added{In CMV, for \punishees that continue to
participate in CMV, the removal {\em causes} a change for the better in
X.}
\end{quote}
\added{We adopt this ``optimistic'' template as a lens on whether moderators can
view their work as not just clean-up (as important as that is), but also as a way
to improve future behavior.  Indeed, given that CMV is both a well-moderated
community and one that stands out for its quality of discussions, it is
sensible to investigate whether the latter feature is a consequence
of the former.
Our concrete hypotheses specify ``change for the better in X'' as follows:} \par
\begin{tabular}{p{5.1in}}
\added{{\bf H-NonCompliance}: decrease in the rate of subsequent rule violations}.
\\
\added{{\bf H-Toxicity}: decrease in use of toxic language.}
\\
\added{{\bf H-Achievement}: increase in contributions achieving community goals (in the case of CMV,
to persuade) and/or receiving community approval.}
\\
\added{{\bf H-Engagement}: increase in level of engagement.}
\end{tabular}

To identify the causal effects of content moderation from observational data,
we could start from a widely adopted method, interrupted time-series analysis (ITS) \cite{james_lopez_2016},
with the removal of a comment serving as the ``interruption''.
However, ITS  does not allow us to discount the following counter-hypothesis: that the observed effects could be explained by the context surrounding the posting of a to-be-removed comment, rather than being caused by the moderation action itself.
Our innovation is to leverage a new quasi-experimental design that we call ``\delayedreaction''; this approach exploits the observation that {\em some \punishees make comments in the (often hours-long) ``\preremoval'' window wherein the 
to-be-removed
 comment has not yet actually been removed}. 
With respect to such \punishees,\footnote{It should be noted these users
constitute
a relatively high-activity
subset of users with respect to the \preremoval window.}
we are able to identify the causal role of comment deletion in reducing
immediate
noncompliance rates,
in comparison to ``control'' users as introduced in \secref{sec:methods}.
This finding supports H-NonCompliance.
For our three other hypotheses, although ITS 
reveals
 some significant changes in
the behavioral aspect of interest before versus after comment removal, 
our \delayedreaction analysis does not attribute these changes to the causal effect of the moderation action.

We conclude our work by discussing implications for content moderation and methodological implications for identifying causal effects from observational data.

\section{Background and Related Work}
\label{sec:related}

In this section, we
first
describe comment removal in CMV
and review prior literature relevant to our four hypotheses.
We then  discuss enforcement and maintenance of social norms in general.

\subsection{Comment Removal in CMV}

\begin{figure}[t]
     \centering
     \includegraphics[width=0.95\textwidth]{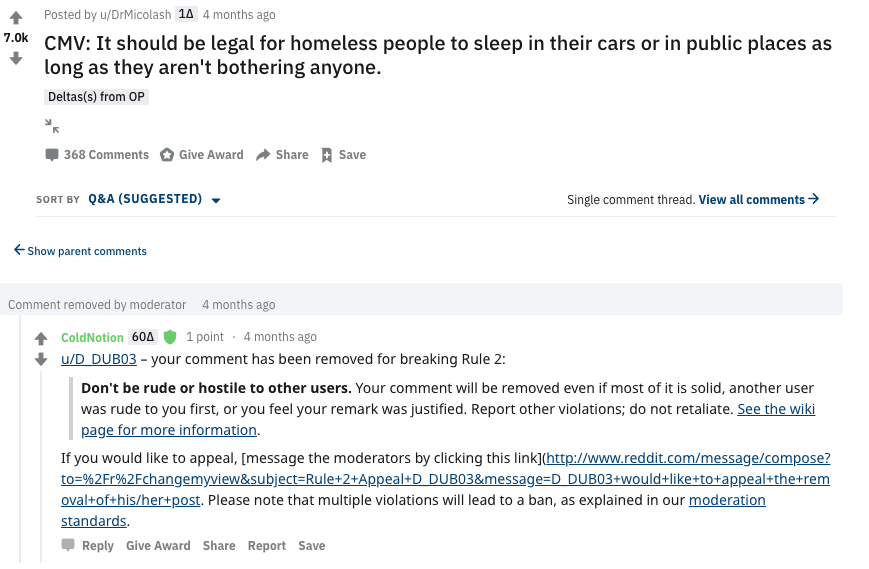}
     \caption{
     Public display for a removed comment.
     A comment by user \textit{D\_DUB03} was removed by moderator \textit{ColdNotion} for violating Rule 2
     and replaced by the text ``Comment removed by moderator''.
     In this case, the moderator replied to the removed comment with the reason for removal and
      also included an explanation of how to appeal.
    Although the comment is no longer visible to the public, its content remains
    available in the moderator interface unless the user deletes it on their end.
     }
     \label{fig:sample_comment}
 \end{figure}

CMV provides a detailed description of
moderation standards and practices through a wiki page that is visible to all users\footnote{\url{https://www.reddit.com/r/changemyview/wiki/modstandards}.}.
A comment may be removed by a moderator if it is considered to violate one or more rules,
such as by being rude or hostile, not challenging the original post,
or violating any Reddit site-wide policy.
Moderators in CMV
tend to be very active in maintaining community standards,
while at the same time
not deleting comments based on topic or content as long as the comments follow the rules.
Removed comments
remain visible to moderators afterwards (unless the author removes it themselves).
As an example, Figure \ref{fig:sample_comment} shows a screenshot of a publicly visible \posttree\footnote{\url{https://www.reddit.com/r/changemyview/comments/9zx07n/cmv_it_should_be_legal_for_homeless_people_to/eadodfn/}.} where moderator \textit{ColdNotion} removed a comment from \textit{D\_DUB03}. The moderator usually replies to the removed comment
and explains how to appeal.

\paragraph{Terminology.} We use the following terminology throughout this paper:
\begin{itemize}
    \item A \introterm{\posttree} is a discussion or conversation tree, rooted at a top-level conversation-starting
original post, and also
containing the replies to the root post, replies to the replies, and so on as nodes.
    \item %
    An \introterm{original poster (OP)} for a \posttree is the author
    of its root.
    \item A \introterm{comment} is a non-root node in a \posttree.
    \item \introterm{\Treatment} {refers to the removal of
    a  comment by a moderator,
    {\em where that removal is the first or second one experienced by the comment's
    author}}, since the third removal
    within six months
    triggers a (temporary) ban.
    Note that the removal rate before the first removal is always 0\%.
    \item An \introterm{\treatmentposttree} {is a \posttree within which a comment removal occurred}.
    \item An \introterm{\treatment user {\rm or an} \punishee} {is the author of a removed comment}.
    \item A \introterm{problematic} comment refers to a comment that would be deemed to violate community rules by moderators, either removed or not.
\end{itemize}

\subsection{H-NonCompliance: Related Work}
\label{sec:noncompliance-prior}
\citet{kiesler+kraut+resnick+kittur:2011} give a thorough consideration of the
tangle of factors at play when it comes to whether or not regulating online behavior
limits the effects or extent of bad behavior (``bad'' is their choice of term).
We enumerate here those of their design principles that relate to
whether we can expect comment removal to reduce the rate of subsequent rule violations
by the authors of removed comments, should they continue to participate in the CMV community.
Kiesler et al.'s (henceforth KKRK) Design Claim 2 states that
authors whose comments are deleted (rather than redirected to some other space)
may post additional inflammatory content;
on the other hand,  CMV comment removals can be appealed,
and thus, by KKRK  Design Claim 3, will be more effective, since users react more
positively to sanctions perceived to be administered fairly. KKRK Design Claim 7 advocates
for a widely-adopted community norm of ignoring (rather than reacting to) trolls,
which perhaps suggests that H-NonCompliance might not hold;
moreover, the usual CMV comment-removal message violates KKRK Design Claim 23's recommendation to allow an
\punishee to save face (e.g., by saying something like ``you might not have known
about this rule'').  On the other hand, since CMV comment removal forms part
of a sequence of more consequential moderation actions (that is, repeated
removals lead to user bans), KKRK Design Claim 31's advocacy of graduated sanctions
suggests support for H-NonCompliance.  In all, we take these contrasting possibilities
as evidence that there is no obvious a-priori answer as to whether H-NonCompliance
holds, and thus it is an interesting hypothesis to explore.

When it comes to the offline world, the theory of specific deterrence --- that a
sanctioned individual will be deterred from the sanctioned behavior in the
future\footnote{{\em General} deterrence concerns whether the behavior of people
other than the \punishees improves.} --- has been shown to be insufficient, with
many modulating factors at play \citep{sherman1993DefianceDeterrence,pratt2006EmpiricalStatus,nagin2018DeterrenceChoice}, such as, as mentioned above, whether the \punishee
perceives the process to be fair. Again, it does not appear possible to draw a
direct prediction regarding H-NonCompliance from this literature.

Yet, inspired in part by (generalized) deterrence theory, \citet{seering_shaping_2017}
hypothesized that banning users that post spam messages on the Twitch
gaming platform would reduce further spamming
rates (by other users). While the spam rate actually increased, it increased
by much less than during a comparable time period in which spammers were {\em not}
banned.  This observation contradicts H-NonCompliance when taken literally,
but can be interpreted as supporting H-NonCompliance in spirit.

 Our work on comment removal on CMV bears strong similarity to
\citeposs{Chang-Recidivism:19} examination of the effect of temporarily banning users on
 Wikipedia, particularly with respect to the chance of future rule violations ---
it goes up, in contradiction of H-NonCompliance, but no causal link is claimed. Their focus differs from ours:
they look at how an affected
user's prior community engagement and perception of the fairness of the block (ban) are
important mediating factors
in post-block behavior.

\subsection{H-Toxicity: Prior Work}

\citet{Chancellor:2016:TIC:2818048.2819963} determine that after Instagram banned
17 tags used to highlight posts that advocate eating disorders,\footnote{Banning tags is a softer form of moderation than comment removal: the content
remains on the site, but searching on such tags bring up no results.
} variant tags arose
among the pro-eating-disorder community that ``depict more vulnerable, toxic, and
`triggering' content'' (pg. 1209) --- ``toxic'' meaning invoking self-loathing
and self-harm --- and the new variants were used more frequently than the old ones.
\citet{Chancellor:2016:TIC:2818048.2819963} indeed recommend against suppressing
content (and hence comment removal), due to negative consequences. Their work does not directly match
H-Toxicity (tag ban vs. comment removal, collective behavior vs. behavior of
the \punishee, Instagram vs. CMV, toxic=harmful vs. toxic=hate speech or
swear words), but may still be taken as indirect evidence against
H-Toxicity.

On the other hand, \citet{chandrasekharan2017you} show that after Reddit banned
several hate-based communities in 2015, participants in those communities that
stayed active on Reddit collectively reduced their hate-speech usage by
``at least 80\%''. Despite the differences with our setting (forum dissolution vs.
comment removal, hate-group subreddits vs. CMV), we can interpret
\citeposs{chandrasekharan2017you} findings as support for H-Toxicity.

\subsection{H-Achievement: Prior Work}

With respect to achievement of community tasks, the first component of
H-Achievement, we mention
\citeposs{cunha:2017:WWM:3041021.3055131}
study of a weight-loss community that finds a causal correlation between
number of support messages and a reported reduction in weight.  We are unaware
of work investigating the effects of negative feedback on accomplishment of
specific goals. (Recall that CMV users strive to change someone's mind, operationally
defined as receiving a ``delta'' badge on a comment.)

With respect to community approval,
\citet{cheng+dnm+leskovec:2014} look at the subsequent reception of commenters'
contributions after these commenters receive negative social feedback, finding
that scores drop, indeed, even more than would seem warranted given computed
text-quality scores (which themselves also decrease). Similarly,
\citet{Ahn:2013:LBQ:2655780.2655784} show that the more downvotes
a Stack Exchange user receives on a question, the lower the score of their subsequent
questions.   Although the settings of these two papers
don't exactly match H-Achievement (online news/Q\&A sites vs. CMV, low community-assigned
scores vs. moderator deletion of comments),  we interpret their results as
contradicting  H-Achievement with respect to its community-approval component.

While a less direct fit to hypothesis H-Achievement, the work of
\citet{cheng+dnm+leskovec:2015} is also related.  One of their experiments
shows that having many comments deleted, as opposed to a few, leads to a drop
in predicted community-approval score.

\subsection{H-Engagement: Prior Work}

\citeposs{cheng+dnm+leskovec:2014} social-feedback experiments, mentioned above,
demonstrate increased commenting rates after receiving low community evaluations,
an observation that supports H-Engagement. However, the temporary-ban study
by  \citet{Chang-Recidivism:19} on Wikipedia found no significant change in future activity rate.

\subsection{
Enforcement and Maintenance of Injunctive Norms: A Broader Landscape
}

The work we present in this paper focuses on a particular moderator reaction
to the violation of explicitly stated community rules.\footnote{We mention a growing line of research on characterizing the rules
themselves in online communities.
\citet{Brian_keegan_2017} study the evolution of rules on Wikipedia by tracking revisions on rule-related Wikipedia pages.
\citet{fiesler+al:2018} provide a characterization of different types of rules and show that community rules share common characteristics across subreddits.
In comparison, \citet{chandrasekharan_internets_2018} perform a large scale study to understand content moderation through the language used in the removed comments on Reddit and identify norms that are universal (macro), shared across certain groups (meso), and specific to individuals (micro).}
Community rules can be viewed as an instantiation of injunctive norms that characterize the perception of what most people approve or disapprove of (vs. descriptive norms that characterize the perception of what most people do) in the focus theory of normative conduct \citep{cialdini1991focus}.
Prior research has studied the effect of {\em exposure} to (violating) such norms on community-level or discussion-level behavior \citep{matias2019preventing,morgan2018welcome,muchnik2013social}.
For instance, through randomizing announcements of community rules to discussions in the Science subreddit,
\citet{matias2019preventing} shows that making community rules visible prevents unruly and harassing conversations.
In comparison, our focus is on the effect of moderator actions that {\em enforce} injunctive norms on individual users.
It is worth remembering that the norms that moderators enforce can change based on circumstances, such as whether humorous content is allowed during a serious situation \citep{leavitt2014upvoting}.

\definecolor{limegreen}{HTML}{32CD32}

\newcommand{\yesyes}{\raisebox{0pt}{\tikz{\node[draw,scale=0.5,diamond,fill=red](){};}}\xspace}
\newcommand{\nono}{\raisebox{0pt}{\tikz{\node[draw,scale=0.5,regular polygon, regular polygon sides=4,fill=limegreen](){};}}\xspace}
\newcommand{\yesno}{\textcolor{blue}{$\CIRCLE$}\xspace}
\newcommand{\probcomm}{\reflectbox{\textcolor{red}{$\lightning$}}\xspace}
\newcommand{\probcommauthor}{$u_\lightning$} %

\section{Methods}
\label{sec:methods}

Our hypotheses are explicitly phrased as causal, positing effects
directly attributable to content moderation. However, we decline, for both
practical and ethical reasons, to use randomized experiments to study these
effects: we do not propose disrupting the functioning of a popular and productive
community, nor do we advocate artificially or randomly altering the selection
of comments to delete.
Instead,
in what follows, we describe quasi-experimental designs --- both existing and new ---
that,
although applied to observational data,
can still provide partial insight into the
causal effects of comment removal.

\begin{figure*}[h]
    \centering
    \includegraphics[width=0.8\textwidth]{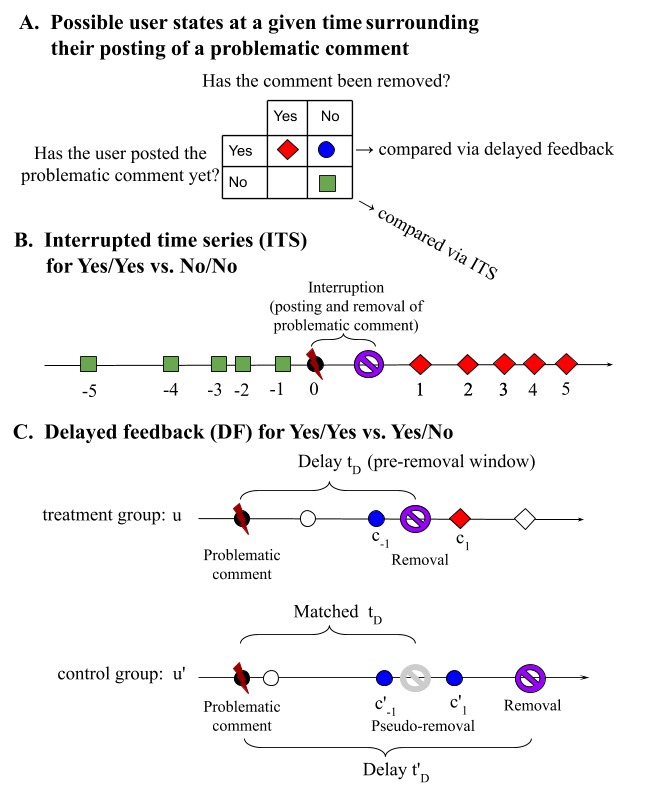}
    \caption{
    \textbf{A.} User situations relevant to our causal hypotheses and how they are compared in our observational setting.
    \textbf{B.} 
    Sketch of our use of interrupted time-series analysis (ITS)
    where the `interruption' consists of the (joint) posting and removal of
    the
    problematic comment.
    \textbf{C.} Sketch of the delayed feedback (DF) design,
    where the `delay' corresponds to the interval between the time of an action (posting of
    the problematic
    comment, in our case) and the time of  `feedback' on that action (its removal, in our case).
    }
    \label{fig:methods}
\end{figure*}

To orient our discussion, Figure ~\ref{fig:methods}A summarizes the set of possible
situations we might ideally want to compare in order to address our
causal hypotheses. These situations consist of possible states a user could be
in around the time that they post a problematic comment.
The top row corresponds to situations in which the user has posted the problematic comment, and either had
this problematic comment removed by the moderators (Yes/Yes) or not (Yes/No).
The bottom row corresponds to situations in which the user has not yet posted the problematic comment, and either had
this problematic comment removed (No/Yes) or not (No/No).
We do note that
the No/Yes case cannot occur by definition, since moderators cannot remove a problematic comment that has not yet been made.%
\footnote{
We assume in this work that moderators do not delete non-problematic comments.
It is beyond the scope of this paper, although potentially fruitful, to
expand our consideration of possible states of interest to those where, say, a user has a
non-problematic comment removed.
}
Moreover, in observational settings, some cases may not turn out to be
directly comparable;
for example, if the moderators never remove a certain type of problematic comment,
then
users who make those problematic comments
will never be part of the data for the Yes/Yes quadrant.
Nevertheless, we can find comparable and accessible sub-cases of
the states as categorized by Figure \ref{fig:methods}A
that allow (a limited) investigation of our causal hypotheses through quasi-experimental designs.

We first (Section \ref{sec:its}) discuss using the standard technique of interrupted time-series analysis,
which, from the perspective of Figure \ref{fig:methods}{A}, compares situations occurring across the main diagonal (Yes/Yes vs. No/No).
This exposes the joint effect on the user's behavior of writing
the problematic comment and
having
the
problematic comment removed.
Second (Section \ref{sec:delayed}), we aim to separate the effect of having
the problematic
comment removed from
the effect
of simply having made a problematic comment
in the first place
--- that is, to see whether a user, after making
the problematic comment,
would behave the same whether or not their problematic comment is removed.\footnote{For example, with respect to H-Toxicity, it could be that when a user posts
a comment with strong swear words, they have usually reached  maximum anger
(or realize they've crossed a line), so their subsequent comments are less
toxic regardless of whether they are officially warned.
}
Towards this aim, we introduce a new \emph{delayed feedback} approach;
this methodology exploits the observation that there is a delay between the moment a problematic comment is posted and the moment it is removed by the moderators, allowing comparisons across Figure \ref{fig:methods}A's top row (Yes/Yes vs. Yes/No).

\subsection{Interrupted Time-Series Analysis: A Standard Design }
\label{sec:its}
In order to observe the (joint) effect of writing a problematic comment and of its removal,
while at the same time controlling for the characteristics of the user,
we employ the standard quasi-experimental design of interrupted time-series  analysis (ITS).
As illustrated in Figure \ref{fig:methods}{B}, this design allows us to  observe and compare between situations that fall across the diagonal of Figure \ref{fig:methods}A.
A given user is in a Yes/Yes situation \emph{after}
the
problematic comment they
posted is removed~(\yesyes); the same user is in a No/No situation
\emph{before} they posted the problematic comment~(\nono). We can then compare the behavior of the user in their first $k$ comments after their problematic comment was removed with that of their last $k$ comments before they posted the problematic comment ($k=5$ in Figure \ref{fig:methods}B).
A discontinuity in behavioral trends that occurs at the time the problematic comment is posted and removed can then be interpreted as being the result of this \emph{interruption}.
Formally, for a given user $u$ we can model the temporal variation of their observed behavior as
\begin{equation}
y(i) = \beta_0 + {\beta_1}i + {\beta_2}x(i) + {\beta_3}(i\cdot{x(i)}),
\label{eq:regression}
\end{equation}
where $i \in \{-k,\ldots, -3, -2, -1, 1, 2, 3, \ldots,k\}$ indexes $u$'s comments
such that $i = -1$ corresponds to the user's last comment before
the `interrupting' problematic comment and $i=1$ corresponds to the user's first comment after the
`interrupting' problematic comment.\footnote{We ignore
the problematic comment itself, and discard cases in which the user made comments between the problematic comment and the time of its removal, in order to avoid ambiguity with respect to the
location
of discontinuity.}
The binary variable $x(i)$ indicates whether the observation is made pre-interruption ($x(i)=0$ for $i<0$)
or post-interruption ($x(i)=1$ for $i>0$).
The pre-interruption slope $\beta_1$ estimates the underlying trend in the absence of an interruption; the level change $\beta_2$ estimates the change in level that can be attributed to the interruption; and the change in slope $\beta_3$ quantifies the difference between the pre-interruption and post-interruption slopes \cite{james_lopez_2016,Kontopantelish2750,Pavalanathan2018MindYP}.
By quantifying discontinuities in pre/post-interruption behavioral
{trends}, this methodology accounts for temporal effects that might otherwise explain a
change between the before and after measurements.

\subsection{Delayed Feedback: A New Design}
\label{sec:delayed}
Despite being a broadly adopted quasi-experimental design, in our setting,
standard
ITS cannot
show that the user, after making a problematic comment, would have behaved
worse if the comment had not been removed --- in which case, one cannot
say that the removal {\em caused} behavior improvement.
For example, a behavioral discontinuity could be
explained by the context surrounding the posting of the problematic comment
(e.g.,
the user's mood might have reached maximum anger when they posted the comment,
so that subsequent comments are less toxic) rather than by the moderation action.
In order to zero in on the effect of
the removal itself,
we need to compare across the top row of Figure \ref{fig:methods}{A} (Yes/Yes vs. Yes/No),
i.e., to take situations where the user has posted a problematic comment, and
compare states wherein the comment has been removed against states where it has not.

Since we can only know that a comment is problematic if it was
removed by a
moderator,\footnote{%
    We do not have the resources to manually check non-deleted comments for problematicity.
} the main difficulty
is to find users
that posted a problematic comment that was not removed (i.e., users in the Yes/No scenario).
Our main intuition, sketched in Figure \ref{fig:methods}{C} (top), is to exploit the fact that moderator actions are not immediate: in fact, in
about 40\% of the cases, it takes more than two hours for the removal to take place (Figure \ref{fig:cdf}).
In particular, behavior observed after the user posted
the
problematic comment but before its removal  --- a delay period which we call the \emph{pre-removal window}, denoting activity within it with a blue circle (\yesno) --- cannot be attributed to the effects of the moderation action that hasn't yet happened.

We could then estimate the effect of removal on the individual that posted
the
problematic comment by comparing their behavior after its removal (comment $c_1$, \yesyes) ---
a time when the user is in a Yes/Yes situation --- with that exhibited
in the pre-removal window (comment $c_{-1}$,\yesno),
wherein even though the user had posted
the
problematic content, they had not yet experienced its removal (Yes/No).
We need, however, to account for the possibility that the observed changes can be attributed to temporal effects (since $c_{-1}$ always occurs before $c_{1}$, and users are expected to change behavior with time).
A discontinuity analysis cannot be used to address this concern as in the ITS approach, due to the limited observation window between the posting and removal of
the problematic
comment.\footnote{On average, a user that has posted at least one comment in the pre-removal window contributed only 0.2 additional comments during that time window in non-affected trees (or 2.5 in affected trees).
\label{footnote:pre-removal-sparse}
}

To account for temporal effects and check whether the change in behavior is indeed due to the moderation action, we rely on a temporally paired
``control'' group (Figure \ref{fig:methods}{C} bottom):
  for each user $u$ that had
  the problematic comment removed with a delay of $t_D$ after the posting time,
   we select a matched user $u'$ that also had a comment removed, but with a slightly larger delay of $t_D'>t_D$.\footnote{Borrowing terminology from randomized experiments into this quasi-experimental design, we will say that $u'$ belongs to the
   ``control'' group and $u$ belongs to the ``treatment'' group.}
We can then consider the difference in  $u'$'s behavior
before and after the pseudo-removal,
i.e., the difference between their behavior in $c_{-1}'$ and in $c_{1}'$ --- to indicate the underlying temporally-driven change that we can expect a user
to undergo
around
$t_D$ seconds
after they made
the
problematic comment.
By discounting this underlying change, we can isolate the effect of comment
removal from that of temporal effects.\footnote{Our choice of control for temporal effects is similar to that used by \citet{oktay2010causal}: they compare change in behavior before vs. after some ``treatment'' event with change in behavior before vs. after a matched pseudo-event.
As an aside, it is interesting to note that
their focus is the reverse of ours, in the sense that the ``treatment'' event is the action itself (in their case, the posting of an answer on a Q\&A site) rather than the feedback this action receives (its selection as a high-quality answer); as such,
they do not
consider the action-to-feedback delay
for matching treatment users with control users.
}

Importantly, this methodology assumes the duration of the delay is not related to the behavior of interest;  while in other domains this was shown to hold true {\cite{seering_shaping_2017}}, this assumption could be challenged in our domain.  To mitigate this risk, we select $u'$ such that their delay time $t'_D$ is as close as possible to $t_D$. Furthermore, the validity of this control can be examined by comparing $c_{-1}'$ and $c_{-1}$, where we would expect no difference in behavior between $u'$ and $u$.

\paragraph{More general applicability} We note that
this delayed-feedback quasi-experimental design can be applied more broadly to estimate the effect of other types of feedback, beyond the removal of a comment.  By exploiting the \emph{delay} between an action and the feedback that action receives,
the approach can disentangle the effect of the feedback from that of the circumstances of the action itself.

\section{Application to the CMV Setting}
\label{sec:churn}

We obtained all\footnote{Throughout, references to ``all'' data are understood to be qualified by the possibility
of inadvertent processing errors, Reddit API issues,
and content deletion by users.
}
73K ChangeMyView \posttrees dating from the subreddit's inception, January 2013, until March 2018 from \url{https://pushshift.io},
a site maintained by Jason Baumgartner.
CMV administrators gave us access to the content and meta-data for
all 23K comments removed by the moderators
during the same period
for breaking CMV's rules.\footnote{Comment removal started in January, 2015, indicating opportunities for community-level natural experiments; pursuit of this idea lies beyond the scope of this paper.}\footnote{A reviewer asked whether CMV notifies users that their deleted comments
    are retained past 30 days.  The CMV wiki entry on moderation standards
    (\url{https://www.reddit.com/r/changemyview/wiki/modstandards}) implies
    that the moderators have access to comment contents past 30 days, since
    otherwise they could not consider appeals of the ``ban when 3 removals have
    occurred within 6 months'' policy.
}
The meta-data contains time of removal, URL,
username,
moderator name, rule violated and description of violation.
Violations range from personal hostility to lack of substance (e.g., pure jokes, nothing but a URL). Table \ref{tab:dataset} provides basic statistics about the dataset.
The public version of our dataset (excluding the content of the removed comments) is available at \url{https://chenhaot.com/papers/content-removal.html},
and will allow other researchers access to our filterings and matchings.

 \begin{table}[t]
  \centering
     \begin{tabular}{p{8cm}r}
         \toprule
        Number of \posttrees  & 73,047\\
        Number of users in CMV who comment at least once & 176,409\\
        Number of comments in CMV  & 4,176,818\\
        Number of comments removed  & 22,788\\
        Number of moderators who removed comments & 43\\
        Number of users in CMV whose comment is removed at least once & 12,481\\
        Total \posttrees with removed comments & 8,463\\
         \bottomrule
     \end{tabular}
     \caption{Statistics of
     our dataset.}
     \label{tab:dataset}
 \end{table}

\begin{figure}[t]
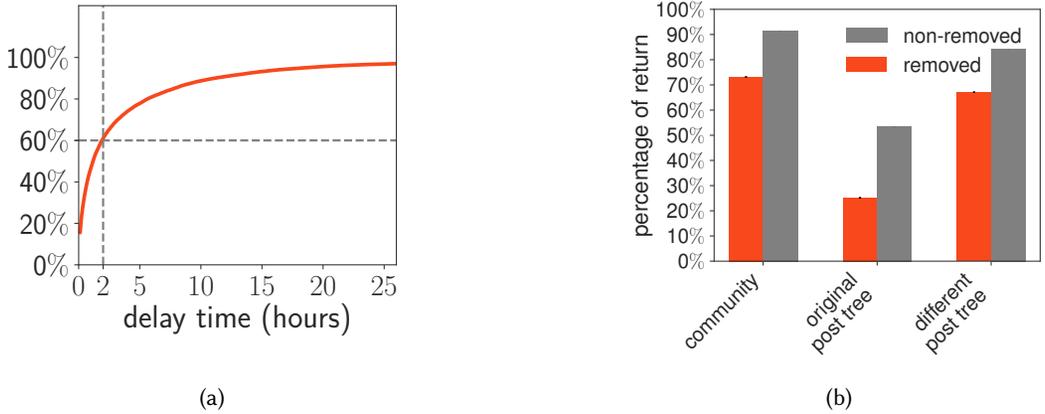

    \centering
    \begin{subfigure}[h]{0.4\textwidth}
        \centering
        \addFigure{\textwidth}{draft/others/cdf_banned_time.pdf}
        \caption{}
\label{fig:cdf}
    \end{subfigure}
    \hfill
   \begin{subfigure}[h]{0.4\textwidth}
        \addFigure{\textwidth}{draft/others/control_treatment_return_percentage_community.pdf}
    \caption{}    \label{fig:community_return_percent}
    \end{subfigure}
    \caption{ (a) Cumulative distribution function of the delay between the time a problematic comment is posted and the time it is removed.
(b)    After making a comment $c$, to what CMV locations, if anywhere, does the author $a$ post afterwards,
    depending on whether $c$ was removed?  \textcolor{red}{Red}: $c$ was $a$'s $1^{\mbox{st}}$ or $2^{\mbox{nd}}$ comment to ever be removed. \textcolor{darkgray}{Gray}/stripes: all other comments.
    The same user can contribute to both colors via different comments.
    ``community'': the union of the original and different \posttrees.
    Error bars represent (tiny) standard errors.
    }
    \label{fig:data_main}
\end{figure}

Our experiments consider only \punishees that continue to participate in CMV after
comment removal.
It is important to point out that
{\em \punishees do abandon CMV at somewhat higher rates than other ChangeMyView members}, as shown in \figref{fig:community_return_percent}.\footnote{This holds even when controlling for prior activity rate.
}
We cannot ascertain whether these community-departing users would have been productive
or disruptive in CMV
had they not incurred a removal.
\Punishees that do return to CMV typically
do not
return to the \treatmentposttree.

\smallskip
\para{Applying the proposed approaches.}
We consider separately effects on the \punishees' behavior in the \treatmentposttree and
the behavior of those same individuals
in \nontreatmentposttrees, as the former is intrinsically tied to the context surrounding the making of the problematic comment, while the latter more broadly affects the community.
Since  interrupted time-series
analysis requires extended observations before and after removal,
as it happens, for data sparsity reasons
we can only apply it to the \nontreatmentposttree scenario.
We are able to apply the \delayedreaction approach in both \nontreatmentposttrees and \treatmentposttrees.
Recall that interrupted time-series analysis allows us to study what happens after comment removal, while
our \delayedreaction design disentangles the effect of comment removal itself from other confounding factors;
we compare user situations surrounding the posting of problematic comments associated with the first or second removal experienced by a user.

For ITS, we consider \punishees who had
10 comments before and after removal in
\nontreatmentposttrees ($k=10$),
excluding the OP themselves and moderators.
We do not include any users who made comments between
posting the problematic comment and the time of its removal
because the discontinuity point
is ambiguous in that case.
This procedure yields 2,752 instances of removals.

To apply the \delayedreaction design to study behavioral changes in \nontreatmentposttrees, we first identify every removal where
the \punishee ($u$)
made at least one comment during
the \preremoval window and at least one comment within one week after removal (both comments are required to be in \nontreatmentposttrees).
For each of these instances, we try to find a matched control instance
as detailed in Section \ref{sec:delayed}, ensuring that
the control \punishee ($u'$) also has at least one comment before
the matched pseudo-removal time, as well as at least one comment between this pseudo-removal time and the time of the actual removal of the problematic comment they posted. We choose matching instances for which the delay after comment time is as close as possible.
Each removal can only be used as a control instance once.
After discarding instances that cannot be matched, we are left with 775 treatment and control instances.
We apply a similar procedure for the \treatmentposttrees scenario by considering only comments that were posted in the \treatmentposttrees, leading to 1,139 treatment and control instances.

\smallskip
\para{Measuring behavioral changes.}
\tableref{tab:features_measured} summarizes all the features
we used to operationalize our hypotheses.
All features are measured based on the properties of a single comment.
In our results figures in \secref{sec:results}, the value depicted at a given
comment index is the average feature value taken over all comments at that index.

\begin{table}[h!]
   \begin{tabular}{p{2.5cm}p{11cm}}
   \toprule
      \textbf{Comment-level features} & \textbf{Description}\\
      \midrule
        \multicolumn{2}{c}{{\bf H-NonCompliance}: decrease in the rate of subsequent rule violations.}\\
        \midrule
            Whether the comment is eventually removed &
        This directly measures
        whether a user
        violates community rules
        (as indicated by a moderator removing a comment). \\
        \midrule
        \multicolumn{2}{c}{{\bf H-Toxicity}: decrease in use of toxic language.}\\
        \midrule
        Swear-word ratio & Fraction of the comment's words that are swear words as defined by LIWC \cite{pennebaker2007linguistic},
        capturing  comment toxicity \cite{Chen:2014,Jiang:2018:Linguistic_Signals}.
        Higher values indicate higher toxicity.\\
            Hate-speech word ratio & Fraction  of the comment's words that are
            hate-speech words as defined by the Hatespeech lexicon \cite{hatespeech,Offensive_Speech_Nithyanand_2017}. \\
        \midrule
        \multicolumn{2}{c}{{\bf H-Achievement}: increase in contributions achieving community goals.}\\
        \midrule
        Whether the comment wins a delta  &
        A ``delta'' is a kind of positive feedback received by a user for changing
        the original poster's view.
        (We ignore deltas awarded by someone other than the OP.)
        Deltas indicate valuable contributions to the CMV community, which is dedicated to changing individuals' views. \\
        Comment score &
        The difference between the number of upvotes and the number of downvotes.
         It represents a measurement of community feedback \citep{tan+lee:15}. \\
         \midrule
          \multicolumn{2}{c}{{\bf H-Engagement}: increase in level of engagement}\\
            \midrule
        Inter-comment time &
        The time gap between a comment and its next comment.
        This feature captures user engagement within the community:
        lower inter-comment time indicates higher engagement \citep{conspiracy_Tanushree_2018,temporal_activity_patterns:andreas2007}.\\
        Word count & Total number of words (excluding stopwords) in the comment, which serves as a proxy for the effort spent on writing a comment. \\
        Depth &
        The depth of the comment as a node in the \posttree, indicating how involved this user was in a conversation.
        Higher values indicate involved, detail-oriented discussions \cite{Choi:2015_Characterizing_Conversation_Patterns,tan2016winning,Discussion_threads_reddit:Weninger2013}.\\
      \bottomrule
    \end{tabular}
     \caption{Features considered in testing our hypotheses regarding user behavior changes after comment removal.
     }
      \label{tab:features_measured}
 \end{table}

\smallskip
\para{Ethical considerations.}
We
acknowledge
that, as discussed in \citet{chandrasekharan_internets_2018},
 the use of removed content from online communities for research purposes is controversial.
Although we were granted moderator access to the content of removed comments,
we will not share any information apart from what is publicly available on Reddit.
(As Figure \ref{fig:sample_comment} shows, the content of the removed comment is
not publicly available.)
Although we have access to the username of
the person making the removed comment (a username which is usually
publicly displayed in the moderator's reply, as in Figure \ref{fig:sample_comment}),
our study does not focus on any single individual.
We thus believe that the potential harm of our study is limited,
and our findings will enable building better rules and governing principles for moderation within the CMV community.

\section{Results}
\label{sec:results}

Next, we present results testing our four hypotheses formulated in the introduction.

\begin{figure*}[h]
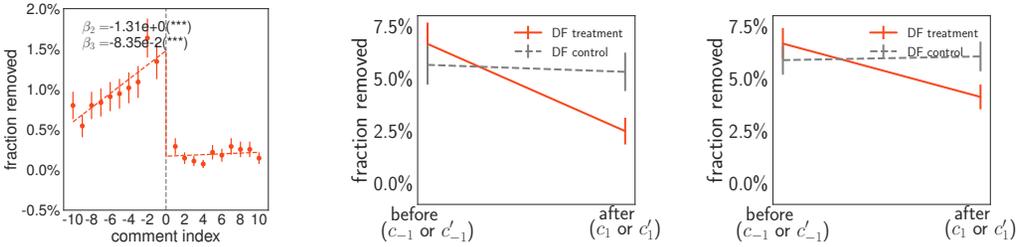

    \centering
    \begin{subfigure}[t]{0.32\textwidth}
    \addFigure{0.85\textwidth}{CSCW/regression_different_post/treatment_first_second_v2_10_percentage_removal.pdf}
    \caption{Interrupted time-series analysis in \nontreatmentposttrees (n=2,752).}
    \label{fig:its_fraction_of_removal}
    \end{subfigure}
    \hfill
    \begin{subfigure}[t]{0.32\textwidth}
        \addFigure{\textwidth}{CSCW/delayed_feedback_one_comment_non_treatment/scaled/percentage_removal.pdf}
    \caption{\Delayedreaction in \nontreatmentposttrees ($\downarrow\downarrow\downarrow$, *, n=775).}
    \label{fig:df_fraction_of_removal_non}
    \end{subfigure}
    \hfill
    \begin{subfigure}[t]{0.32\textwidth}
        \addFigure{\textwidth}{CSCW/delayed_feedback_one_comment_treatment/scaled/percentage_removal.pdf}
    \caption{\Delayedreaction in \treatmentposttrees ($\downarrow\downarrow$, *, n=1,139).}
    \label{fig:df_fraction_of_removal_treatment}
    \end{subfigure}
    \caption{We use the following visualization scheme throughout the paper.
    In interrupted time-series analysis, $*s$ in the parentheses (if any) indicate $p$-values for significant changes via $t$-tests in the regression (*: $p <$ 0.05, **: $p <$ 0.01, ***: $p <$ 0.001).
    Our hypotheses can be supported by significant changes in either $\beta_2$ or $\beta_3$.
    In the delayed feedback approach, we use a $t$-test for evaluating whether user behavioral features
    in the first comment after removal ($c_1$) are significantly different from
    the last comment ($c_{-1}$) in the \preremoval window for the treatment group.
    Up-arrows indicate significance level of an {\em increase} in feature value; $\uparrow\uparrow\uparrow$:  $p <$ 0.001, $\uparrow\uparrow$: $p <$ 0.01; $\uparrow$: $p <$ 0.05. Down-arrows show the same except for a {\em decrease} of feature value.
    We also use $*s$ to indicate whether the difference in difference ($(c_1 - c_{-1}) - (c'_1 - c'_{-1})$) is statistically significant (*: $p <$ 0.05, **: $p <$ 0.01, ***: $p <$ 0.001).
    Error bars indicate standard errors.\\[0.5em]
    {\bf H-NonCompliance:} ITS shows that noncompliance rate decreases after comment removal, and delayed feedback
    identifies the causal role of comment removal in reducing noncompliance rate
    both in \nontreatmentposttrees and \treatmentposttrees.
    }
    \label{fig:noncompliance}
\end{figure*}

\subsection{H-NonCompliance (\figref{fig:noncompliance})}
\label{sec:noncompliance}

Our first hypothesis is concerned with the change in the rate of rule violations ({\em noncompliance rate}), measured by the fraction of comments removed.
Ideally, moderators wish that noncompliance rate decreases after comment removal because comment removal is meant to serve as a warning to help members regulate their behavior rather than escalate the situation and lead to backfire effects.

Indeed, we observe a significant decrease in noncompliance rate after comment removal via interrupted time-series analysis, both for level and slope.
In fact, the fraction of comments removed keeps going up before comment removal, which indicates that in addition to the comment removal performed by moderators, another plausible explanation for reduced noncompliance is due to reaching a local maximum in violating rules and \punishees deciding to reset themselves.

Our \delayedreaction approach allows us to zero in on the effect of comment removal itself.
We find that noncompliance rate declines in the comment right after removal compared to the comment right before removal, both in \nontreatmentposttrees and \treatmentposttrees.
In comparison, we do not observe a similar decrease in the control group.
The fact that the treatment group and the control group
have similar noncompliance rates before removal
supports the validity of our matching scheme.
Overall, these observations confirm the causal role of comment removal in reducing noncompliance rates,
echoing \citet{seering_shaping_2017} in spirit.

It is also important to note that the noncompliance rate in the delayed feedback approach is much higher than that in interrupted time-series analysis,
which indicates that \punishees that remained active in the \preremoval window may be generally more noncompliant than average \punishees.
Within the delayed feedback approach, the noncompliance rate is greater in \treatmentposttrees than in \nontreatmentposttrees, especially after comment removal.
This suggests that comment removal is more effective in regulating user behavior outside the situation that leads to the making of a removal-incurring comment.

\begin{figure*}[t]
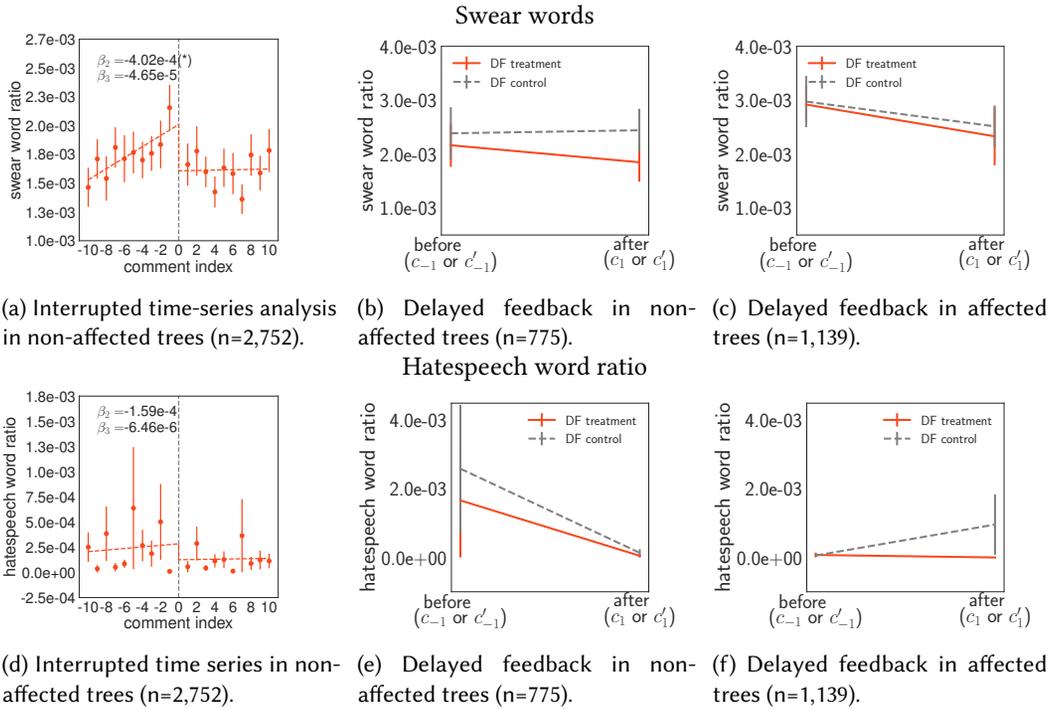

    \centering
    Swear words \\
    \begin{subfigure}[t]{0.32\textwidth}
    \addFigure{0.85\textwidth}{CSCW/regression_different_post/treatment_first_second_v2_10_swear.pdf}
    \caption{Interrupted time-series analysis in \nontreatmentposttrees (n=2,752).}
    \label{fig:its_swear}
    \end{subfigure}
    \hfill
    \begin{subfigure}[t]{0.32\textwidth}
        \addFigure{\textwidth}{CSCW/delayed_feedback_one_comment_non_treatment/scaled/swear_word_ratio.pdf}
    \caption{\Delayedreaction in \nontreatmentposttrees (n=775).}
    \label{fig:df_swear_non}
    \end{subfigure}
    \hfill
    \begin{subfigure}[t]{0.32\textwidth}
        \addFigure{\textwidth}{CSCW/delayed_feedback_one_comment_treatment/scaled/swear_word_ratio.pdf}
    \caption{\Delayedreaction in \treatmentposttrees (n=1,139).}
    \label{fig:df_swear_treatment}
    \end{subfigure}
    Hatespeech word ratio\\
    \begin{subfigure}[t]{0.32\textwidth}
    \addFigure{0.85\textwidth}{CSCW/regression_different_post/treatment_first_second_v2_10_hate.pdf}
    \caption{Interrupted time series in \nontreatmentposttrees (n=2,752).}
    \label{fig:its_hate}
    \end{subfigure}
    \hfill
    \begin{subfigure}[t]{0.32\textwidth}
        \addFigure{\textwidth}{CSCW/delayed_feedback_one_comment_non_treatment/scaled/hate_speech_ratio.pdf}
    \caption{\Delayedreaction in \nontreatmentposttrees (n=775).}
    \label{fig:df_hate_non}
    \end{subfigure}
    \hfill
    \begin{subfigure}[t]{0.32\textwidth}
        \addFigure{\textwidth}{CSCW/delayed_feedback_one_comment_treatment/scaled/hate_speech_ratio.pdf}
    \caption{\Delayedreaction in \treatmentposttrees (n=1,139).}
    \label{fig:df_hate_treatment}
    \end{subfigure}
    \caption{{\bf H-Toxicity:} Use of swear words significantly decreases after comment removal in ITS,
    but we do not observe any statistically change in the \delayedreaction approach.
    None of the hate-speech changes are statistically significant, perhaps
    because the incidence rate is so low to begin with.
    }
    \label{fig:toxicity}
\end{figure*}

\subsection{H-Toxicity (\figref{fig:toxicity})}
\label{sec:toxicity}

Our second hypothesis is concerned with 
use of toxic language, another bad behavior that moderators may wish to discourage via comment removal.
We focus on the use of swear words in this section, because
hate speech turns out to be rarely used by \punishees and none of the changes
turn out to be statistically significant.

Interrupted time-series analysis shows a decrease in the use of swear words,
the level change ($\beta_2$) (only) is statistically significant.
However, this behavior change may not be caused by comment removal, because
the \delayedreaction approach shows no statistically significant change in \nontreatmentposttrees or in \treatmentposttrees, although there is a declining trend for the treatment group in both settings.
Again, the fact that the treatment group and the control group are well matched in swear word ratio before removal supports the validity of our matching scheme.
Overall, it seems that decreases in toxic-language use may not be induced by comment
removal, at least for the comment immediately following comment removal.

\begin{figure*}[h]
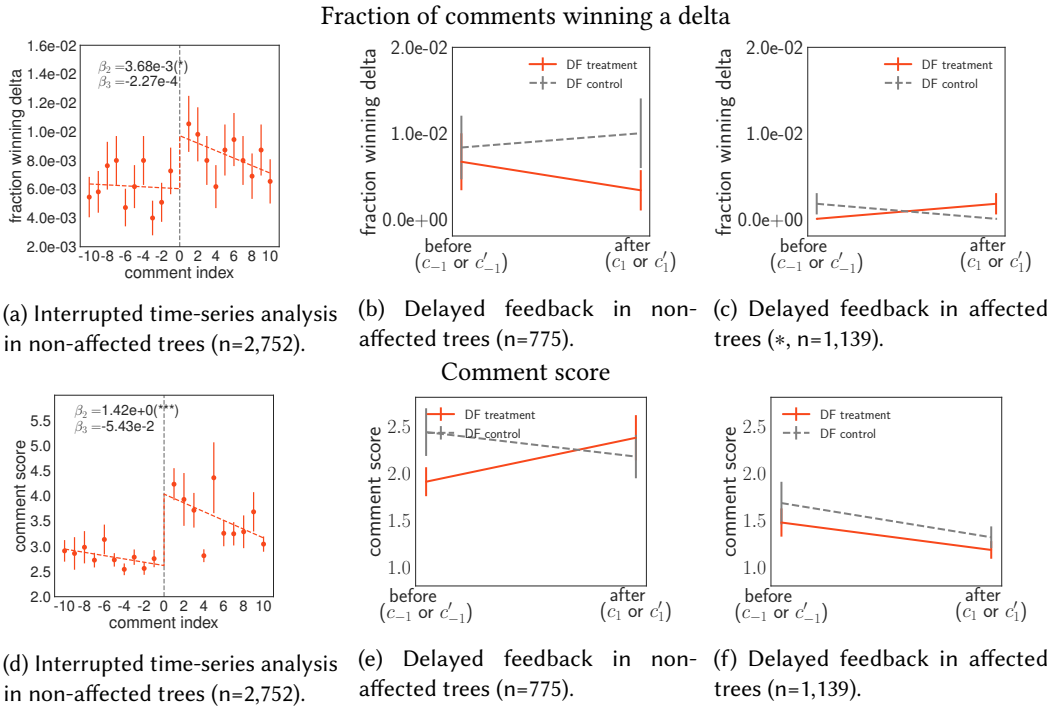

    \centering
    Fraction of comments winning a delta \\
    \begin{subfigure}[h]{0.32\textwidth}
    \addFigure{0.85\textwidth}{CSCW/regression_different_post/treatment_first_second_v2_10_delta.pdf}
    \caption{Interrupted time-series analysis in \nontreatmentposttrees (n=2,752).}
    \label{fig:its_delta}
    \end{subfigure}
    \hfill
    \begin{subfigure}[h]{0.32\textwidth}
        \addFigure{\textwidth}{CSCW/delayed_feedback_one_comment_non_treatment/scaled/delta.pdf}
    \caption{\Delayedreaction in \nontreatmentposttrees (n=775).}
    \label{fig:df_delta_non}
    \end{subfigure}
    \hfill
    \begin{subfigure}[h]{0.32\textwidth}
        \addFigure{\textwidth}{CSCW/delayed_feedback_one_comment_treatment/scaled/delta.pdf}
    \caption{\Delayedreaction in \treatmentposttrees ($*$, n=1,139).}
    \label{fig:df_delta_treatment}
    \end{subfigure}\\
    Comment score\\
    \begin{subfigure}[h]{0.32\textwidth}
    \addFigure{0.85\textwidth}{CSCW/regression_different_post/treatment_first_second_v2_10_mean_score.pdf}
    \caption{Interrupted time-series analysis in \nontreatmentposttrees (n=2,752).}
    \label{fig:its_score}
    \end{subfigure}
    \hfill
    \begin{subfigure}[h]{0.32\textwidth}
        \addFigure{\textwidth}{CSCW/delayed_feedback_one_comment_non_treatment/scaled/mean_score.pdf}
    \caption{\Delayedreaction in \nontreatmentposttrees (n=775).}
    \label{fig:df_score_non}
    \end{subfigure}
    \hfill
    \begin{subfigure}[h]{0.32\textwidth}
        \addFigure{\textwidth}{CSCW/delayed_feedback_one_comment_treatment/scaled/mean_score.pdf}
    \caption{\Delayedreaction in \treatmentposttrees (n=1,139).}
    \label{fig:df_score_treatment}
    \end{subfigure}
    \caption{{\bf H-Achievement:} Although ITS shows significant increases in user achievement both in winning a delta and comment score, there is no significant change in any settings with the \delayedreaction approach.
    }
    \label{fig:achievement}
\end{figure*}

\subsection{H-Achievement (\figref{fig:achievement})}
\label{sec:achievement}

Our third hypothesis examines the good behavior that members can contribute to CMV, including making comments that actually change other users' views or that receive positive community feedback as measured by the difference between the number of upvotes and number of downvotes that a comment receives.
Given that comment removal is a strategy for mitigating bad behavior, any positive effect on future good behavior would be notable for moderators.

Using interrupted time-series analysis, we observe a positive level change both in fraction of comments winning a delta and comment score, which suggests potential positive effects of comment removal.
However, we cannot confirm the causal role of comment removal with our \delayedreaction approach.
Although the difference in difference is statistically significant in \treatmentposttrees for winning a delta (\figref{fig:df_delta_treatment}), the treatment group and the control group are not well matched in the \preremoval window.
The trends are also mixed for different combinations of achievement measures and \nontreatment/\treatmentposttrees.
This contrast suggests that the improvements in achievement might be due to temporal effects, such as getting out of the situation that leads to the making of a removal-incurring comment.
Note that fraction of comments winning a delta in \treatmentposttrees and comment score in \nontreatmentposttrees are the only cases where the treatment group and the control group are not well matched in the \preremoval window.

\begin{figure*}[h]
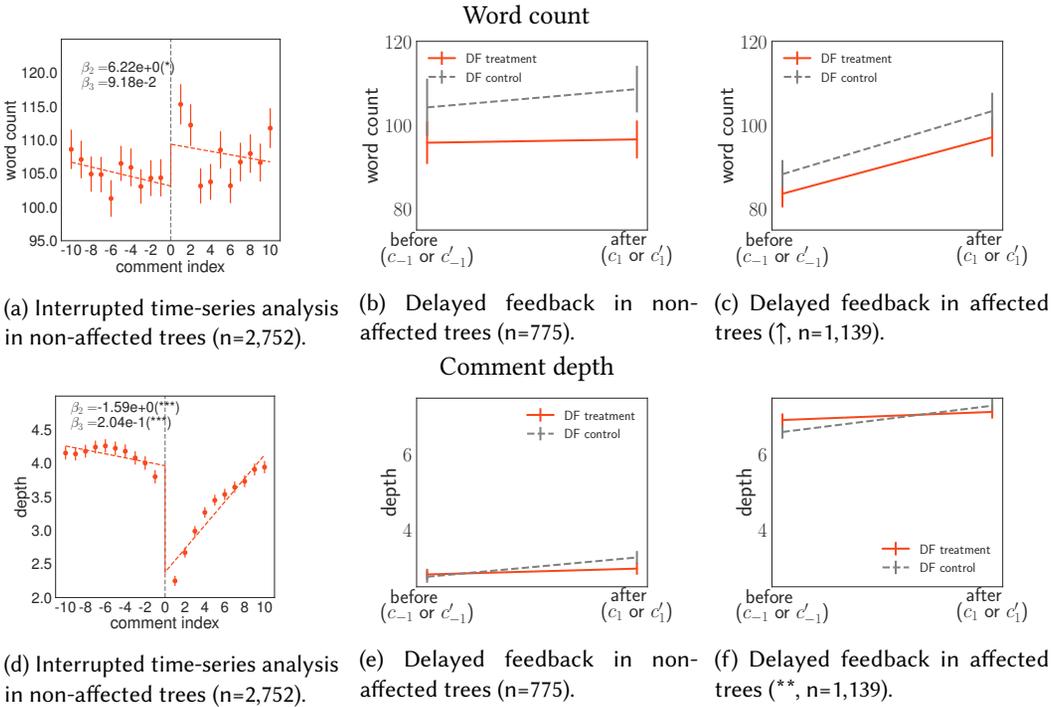

    \centering
    Word count\\
    \begin{subfigure}[h]{0.32\textwidth}
    \addFigure{0.85\textwidth}{CSCW/regression_different_post/treatment_first_second_v2_10_mean_wc.pdf}
    \caption{Interrupted time-series analysis in \nontreatmentposttrees (n=2,752).}
    \label{fig:its_wc}
    \end{subfigure}
    \hfill
    \begin{subfigure}[h]{0.32\textwidth}
        \addFigure{\textwidth}{CSCW/delayed_feedback_one_comment_non_treatment/scaled/mean_word_count.pdf}
    \caption{\Delayedreaction in \nontreatmentposttrees (n=775).}
    \label{fig:df_wc_non}
    \end{subfigure}
    \hfill
    \begin{subfigure}[h]{0.32\textwidth}
        \addFigure{\textwidth}{CSCW/delayed_feedback_one_comment_treatment/scaled/mean_word_count.pdf}
    \caption{\Delayedreaction in \treatmentposttrees ($\uparrow$, n=1,139).}
    \label{fig:df_wc_treatment}
    \end{subfigure}
    Comment depth\\
    \begin{subfigure}[h]{0.32\textwidth}
    \addFigure{0.85\textwidth}{CSCW/regression_different_post/treatment_first_second_v2_10_mean_depth.pdf}
    \caption{Interrupted time-series analysis in \nontreatmentposttrees (n=2,752).}
    \label{fig:its_depth}
    \end{subfigure}
    \hfill
    \begin{subfigure}[h]{0.32\textwidth}
        \addFigure{\textwidth}{CSCW/delayed_feedback_one_comment_non_treatment/scaled/depth.pdf}
    \caption{\Delayedreaction in \nontreatmentposttrees (n=775).}
    \label{fig:df_depth_non}
    \end{subfigure}
    \hfill
    \begin{subfigure}[h]{0.32\textwidth}
        \addFigure{\textwidth}{CSCW/delayed_feedback_one_comment_treatment/scaled/depth.pdf}
    \caption{\Delayedreaction in \treatmentposttrees (**, n=1,139).}
    \label{fig:df_depth_treatment}
    \end{subfigure}
    \caption{{\bf H-Engagement:}
    ITS shows a statistically significant level change in word count, and statistically significant level and slope changes for comment depth.
    In the \delayedreaction approach, although the change in word count is  statistically significant in the treatment group in \treatmentposttrees, the control group also shows a significant change,
    which suggests that the change in the treatment group is likely due to a temporal effect.
    }
    \label{fig:engagement}
\end{figure*}

\begin{figure}[h]
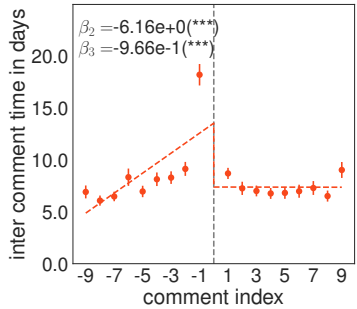

    \addFigure{0.35\textwidth}{CSCW/regression_different_post/treatment_first_second_v2_10_inter_comment_time.pdf}
    \caption{{\bf H-Engagement:} Interrupted time-series analysis shows that comment frequency significantly increases (inter-comment time decreases) after comment removal (n=2,752).  %
     Our \delayedreaction approach cannot be applied to comment rate.}

    \label{fig:inter_comment_time}
\end{figure}

\subsection{H-Engagement (\figref{fig:engagement} and \figref{fig:inter_comment_time})}
\label{sec:engagement}

Our final hypothesis considers changes in level of engagement, such as posting more frequently and writing longer comments.  Communities generally try to foster user engagement,
although we caution that engagement should be considered more a side benefit than
necessarily a goal in moderation on its own.

Using ITS, we find that word count increases after comment removal (statistically significant level change), and depth significantly changes after comment removal (a decline in the level and an increase in the slope).
This suggests that in \nontreatmentposttrees, users move to discussions of shallow depth and then engage deeply again.
In comparison, we do not observe significant changes in \nontreatmentposttrees with the delayed feedback approach.

Note that while  the change in word count for the treatment group is statistically significant in \treatmentposttrees,
the control group also shows a significant increase in word count, suggesting that this shift can be attributed to a temporal effect.
For example, it could be the state of the discussion in which the treatment and control group users were participating,  rather than comment removal, that leads to users making longer comments.
This observation highlights the importance of controlling for temporal effects in the delayed feedback approach.

Although the change in depth for the treatment group is not statistically significant, the treatment group does not dive into deeper discussions in \treatmentposttrees compared to the control group, 
in a statistically significant way.

Finally, using ITS, we find that comment frequency significantly increases after comment removal, both in the level and in the slope.
While this observation is consistent with \citet{cheng+dnm+leskovec:2014}, we cannot claim that this change is due to comment removal, because we cannot apply the delayed feedback approach with this metric (inter-comment time is ill-defined given only one comment).
\section{Concluding Discussion}
\label{sec:conclusion}

We have presented
a systematic study of user behavior changes after comment removal,
using the ChangeMyView community as a testbed.
Our main contribution is to employ multiple methodologies
to attempt to identify the causal role of the comment-removal moderation action.
Specifically, in addition to the commonly-used interrupted time-series analysis, we propose a new \delayedreaction quasi-experimental design to disentangle the effect of comment removal from that of the context
surrounding the problematic comment being posted in the first place.
\tableref{tab:summary} shows a summary of our key results.
Interrupted time-series analysis shows that
\punishees not only regulate their bad behavior
after comment removal, as
reflected by reduced noncompliance and reduced toxicity in language,
but also participate at increased rates and with increased achievement.
Our \delayedreaction approach, however, identifies a \emph{causal} effect of comment removal only in reducing future noncompliance rates, not in inducing other behavioral improvements.

\begin{table}[t]
  \small
  \centering
      \begin{tabular}{p{0.18\textwidth}p{0.24\textwidth}p{0.24\textwidth}p{0.24\textwidth}}
   \toprule
      Hypothesis & ITS in \nontreatmentposttrees: users with 10 comments before and after comment removal, and no comments in the \preremoval window, $n=2,752$ & DF in \nontreatmentposttrees: users that posted in the \preremoval window and after removal in \nontreatmentposttrees, and can be matched, $n=755$ & DF in \treatmentposttrees: users that posted in the \preremoval window and after removal in \treatmentposttrees, and can be matched, $n=1,139$\\
      \midrule
      H-NonCompliance &\multicolumn{1}{c}{\checkmark} & \multicolumn{1}{c}{\checkmark} & \multicolumn{1}{c}{\checkmark}\\
      H-Toxicity &\multicolumn{1}{c}{\checkmark} & & \\
      H-Achievement  &\multicolumn{1}{c}{\checkmark} & & \\
      H-Engagement & \multicolumn{1}{c}{\checkmark} & & \\
      \bottomrule
    \end{tabular}
     \caption{Summary of our key results.}
      \label{tab:summary}
 \end{table}

\smallskip
\para{Implications for content moderation.}
Our results show that comment removal as a content moderation strategy is successful in regulating
one type of bad behavior
and potentially useful for encouraging good behavior in the CMV community,
at least in the short term,
since
it consistently results
  in reduced noncompliance both when analyzed via interrupted time-series regression and when analyzed with our delayed feedback approach.
Delayed feedback shows the results holding in both
 \nontreatmentposttrees and \treatmentposttrees, which suggests that comment removal does not
 (immediately)
 backfire even in the situation that leads to the making of a removal-incurring comment.

It is important to note that ITS and delayed feedback capture disjoint samples.
The fact that noncompliance decreases for two disjoint samples further validates the effectiveness of comment removal
in reducing rule violation in CMV.\footnote{It is worth noting that in an early version of our delayed feedback approach, we observed an
eventual
increase in noncompliance rate when controls were not applied.
}

Furthermore,  ITS reveals other positive behavior changes after removal,
specifically, improvements in
language non-toxicity, user achievement, and user engagement.
However,
since ITS does not account for the context of the problematic comment, we cannot causally attribute these effects to the moderation action.

Following the regulatory design discussions in \citet{kiesler+kraut+resnick+kittur:2011}, we see productive work ahead in {exploring the following questions, and their corresponding design implications.}
Given our observation that noncompliance rate is much higher in \treatmentposttrees than \nontreatmentposttrees,
should certain users be temporarily banned from contributing to the \treatmentposttree,
rather than just warned via comment removal?
For the \punishees that are so discouraged by comment removal as to never
return to the site,
is there some way we could identify those who could potentially have been productive
members if they had remained, and can we improve the return rate for such users
by delaying the removal notification?
Is it possible to predict diverse subsequent behavior of \punishees and give them tailored moderation feedback?
In our preliminary experiments, we find that it is feasible to forecast whether an \punishee will make another problematic comment  based on user-level information,
and linguistic cues have also been shown to signal future misbehavior \cite{zhang2018conversations,liu_forecasting_2018,Chang-Trouble:19}.
Such predictive tools could, in the future, assist human moderators in their activities .

The latter two of the three previous questions also suggest a trade-off between the potential positive effect of enforcing community rules and the potential negative effect of driving people away with overzealous moderation.
It is important to develop novel methods for understanding to what extent content moderation is necessary.

Finally, how does comment removal impact other users involved in the discussion?
Although we have focused on users that are directly affected by comment removal in this work,
it is important to understand the potential for spillover effects on the entire community.
We are excited about the prospect of future research into these questions
surrounding tools for algorithmically-assisted moderation.
Meanwhile, we caution against automatic content moderation given the inherent biases and potentially complex consequences \cite{park_reducing_2018,wiegand_detection_2019,sap_risk_2019}.

\smallskip
\para{Methodological implications.}
Our study
engages with
methodological challenges in using observational data to understand
the effects of content moderation.
In the case of problematic-comment moderation,
one should distinguish between the ``interruption'' comprised of
the posting of the comment,  the ``interruption'' comprised of its removal,
and the ``interruption'' comprised of the passing of time.
In order to disentangle the effect of comment removal from the
circumstances surrounding the problematic comment's posting,
we propose a delayed feedback approach that (i) takes advantage of the fact that users may remain active
in the interval between their posting of a problematic comment and its removal by community moderators,
and (ii)
controls for the first and third ``interruptions'' just mentioned  by
finding users with a slightly longer delay between problematic-comment creation
and removal.\footnote{Recall from Section \ref{sec:delayed} and the text surrounding
footnote \ref{footnote:pre-removal-sparse}
that we cannot just apply ITS to the pre-removal vs. post-removal windows, due
to data sparseness issues.}
It turns out that this control group is useful for filtering spurious results, e.g., a
``finding'' of an
 increase in word count after comment deletion.

It is important to note that these methodological choices can affect characteristics of the samples that we end up analyzing.
This sample selection effect can limit the generalizability of our findings.
Moreover, sample-size considerations limit us, in the \delayedreaction approach, to examining a single comment before and after removal, which means we can only capture \punisheespos immediate behavior changes.

Nevertheless, overall, we believe that delayed feedback represents a promising direction for future work in understanding the effect of
content moderation, because
a time gap always exists before manual content moderation occurs.
As a community, we need to further develop observational methodologies for understanding important questions when ethical concerns prevent randomized experiments.

\smallskip
\para{Limitations based on choice of community.}
The positive effect of comment removal could relate to the high-quality moderation of CMV and the commitment of the CMV community.
In particular, the almost universally positive results with interrupted time-series analysis could relate to the fact that users with at least 20 comments in CMV are committed to the goals of CMV and are willing to reflect and make changes for the betterment of this community.
It is also possible that comment removal in CMV applies a consistent standard (thus perceived as fair) that other communities may not emulate.
Finally, CMV is a task-driven community with a clear goal of changing others' views and providing counterarguments.
This goal may help moderators and members orient themselves, but
other communities, such as news discussion forums, are not as goal-centered.

Examining the generalizability of our results to other online communities requires further work.
If we observe different outcomes, it is useful to develop methods of characterizing the practice of moderation,
 such as how
 consistently moderators are able to apply the rules, and understand whether the type of moderation practice affects the outcomes.

\smallskip
\para{Limitations based on data requirements of our methodology.} As we discussed throughout the paper, selection bias is an important limitation of our experiments based on observational data.
We could only run our
analyses on users exhibiting sufficiently high levels of certain types of activity:
we can make no claims regarding \punishees that did not make ``enough'' comments
in the targeted locations before and after the removal happened, including those that choose to leave the community after comment removal.
Furthermore, given the anonymous nature of Reddit, it is possible that users switch to a different account after comment removal or being banned from the community; our approaches do not attempt to link accounts by the same user.
It follows that our findings only apply to those users that remain active with the same account, and the positive behavior changes may not apply to people who choose to switch accounts.
Another limitation is that due to the small sample size, we analyzed all comment removals in aggregate, without differentiating the reasons for removal.

Finally, we note that
  our study was not based on pre-established hypotheses, operationalization, and analysis plans,
in the sense that our initial set of hypotheses have not heretofore been reported in this paper.\footnote{Our initial hypotheses were of the form, ``comment removal causes a change in subsequent behavior'', where
5 more characterizations of behavior
(rates of hedging, type-to-token ratio, positive words, negative words, questions)
were initially considered beyond the 8 reported in Table \ref{tab:features_measured}; following the reviewers' suggestions, we re-organized our hypotheses and in the process dropped those that did not fit well with this re-organization.}
As such, our study should be considered exploratory.
Nevertheless, our findings and delayed feedback methodology can inform the design of experiments that can further probe the causal nature of
moderation
 actions.

\bibliographystyle{ACM-Reference-Format}
\bibliography{refs}
\end{document}